\documentclass[useAMS]{mn2e}

\usepackage{times}
\usepackage{amssymb}
\usepackage{rotating}
\usepackage{graphicx,epstopdf}
\usepackage{longtable}
\usepackage{array}
\usepackage{multirow}
\usepackage{tabularx}
\usepackage{caption}
\usepackage{amsmath} 
\usepackage{color}
\usepackage{siunitx}
\usepackage{float}
\usepackage[dvipsnames]{xcolor}
\usepackage{tabularx}
\usepackage{colortbl}
\usepackage{gensymb}
\usepackage{undertilde}

\newcommand{\thickhline}{%
    \noalign {\ifnum 0=`}\fi \hrule height 1pt
    \futurelet \reserved@a \@xhline
}

\newcommand{\pf}{\textsc{propfluc}}
\newcommand{\ro}{$r_{\text{o}}$}
\newcommand{\ri}{$r_{\text{i}}$}

\newcommand{\rbw}{$r_{\text{bw}}$}
\newcommand{\rg}{R$_{\text{g}}$}

\title[Modelling Cygnus X-1 with propagating fluctuations]{Modelling hard and soft state of Cygnus X-1 with propagating mass accretion rate fluctuations}
\author[S. Rapisarda, A. Ingram and M. van der Klis]{S. Rapisarda$^1$, A. Ingram$^1$ and M. van der Klis$^1$
\\
$^{1}$Anton Pannekoek Institute for Astronomy, University of Amsterdam, Science Park 904, 1098XH Amsterdam, Netherlands\\
}
\date{accepted for publication in MNRAS}
\pagerange{\pageref{firstpage}--\pageref{lastpage}} \pubyear{2017}

\begin{document}


\pagerange{\pageref{firstpage}--\pageref{lastpage}} \pubyear{2017}

\maketitle
\topmargin = -0.5cm
\label{firstpage}

\begin{abstract}
We present a timing analysis of three \textit{Rossi X-ray Timing Explorer} observations of the black hole binary Cygnus X-1 with the propagating mass accretion rate fluctuations model \pf{}. The model simultaneously predicts power spectra, time lags, and coherence of the variability as a function of energy. The observations cover the soft and the hard state of the source, and the transition between the two. We find good agreement between model predictions and data in the hard and in the soft state. Our analysis suggests that in the soft state the fluctuations propagate in an optically thin hot flow extending up to large radii above and below a stable optically thick disc. In the hard state, our results are consistent with a truncated disc geometry, where the hot flow extends radially inside the inner radius of the disc. In the transition from soft to hard state, the characteristics of the rapid variability are too complex to be successfully described with \pf{}. The surface density profile of the hot flow predicted by our model and the lack of QPOs in the soft and hard state, suggest that the spin of the black hole is aligned with the inner accretion disc and therefore probably with the rotational axis of the binary system.  
\end{abstract}


\section{Introduction}
\label{sec:int}

Up to date, the propagating mass accretion rate fluctuations model (Lyubarskii 1997; Kotov et al. 2001; Arevalo \& Uttley 2006, hereafter AU06; Ingram \& van der Klis 2013, hereafter IK13) is one of the most promising models for quantitatively explaining the rapid variability observed in Black Hole X-ray Binaries (BHBs). In the Low-Hard State (LHS), when the energy spectrum is dominated by a power law ($\Gamma \approx$ 1.5) and the luminosity is low, the rapid variability is characterized by broad band noise (fractional rms $\approx$ 30\%) and by a strong quasi-periodic oscillation, or QPO (e.g. Remillard \& McClintock 2006; Belloni 2010; Gilfanov 2010). The shape of the power spectrum can be quite complex, showing several broad band continuum components or ``humps'' in the frequency domain. The study of this variability in different energy bands revealed that hard X-ray variations are often delayed with respect to soft X-ray variations. The amplitude of this \textit{hard lag} correlates with energy and it is larger for variations on longer time-scales (e.g. Miyamoto \& Kitamoto 1989; Nowak et al. 1999). In the propagating fluctuations scenario, broad band noise arises because of mass accretion rate fluctuations stirred up at each radius of the accretion flow and propagating towards the black hole (BH). If the spectrum of the region close to the BH is harder than that emitted at larger radii, the propagation of fluctuations naturally produces hard phase lags (Kotov et al. 2001; AU06). Propagating fluctuations also explain other observational characteristics of BHBs: the linear rms-flux relation on different time scales (Uttley \& McHardy 2001), the high coherence of the variability across a broad range of energy bands (Vaughan \& Nowak 1997; Nowak et al. 1999), and the large amplitude of the X-ray variations (of tens of per cent fractional rms) observed over several decades of time-scales (e.g. Reig, Papadakis \& Kylafis 2002).\\
Kotov et al. (2001) and AU06 made a first step in applying the propagating fluctuations paradigm quantitatively to BHBs. They showed that the model can predict the observed ratio between power in broad soft and hard bands, and the phase lag between these two energy bands for some selected Cygnus X-1 observations. The model \pf{} (IK13; Rapisarda et al. 2016, hereafter RIKK16; Rapisarda et al. 2017, hereafter RIK17) further explores the propagating fluctuations model predictions. \pf{} can produce multi-hump power spectra in different energy bands assuming fluctuations stirred up in and propagating from a truncated disc (optically thick and geometrically thin) through a hot flow (optically thin and geometrically thick). This means the model also predicts the frequency-dependent phase lag between soft and hard energy band. Supplementing propagating fluctuations with solid-body precession of the entire hot flow, the model produces a QPO on top of the broad band noise. The model predicts amplitude, phase lags, and coherence of the rapid variability in and between the soft and hard bands. This represents the complete information that can be extracted by first- and second-order Fourier analysis from the variability in two energy bands (without considering higher order cumulants: the bi-spectrum, tri-spectrum, etc).\\
Rapisarda et al. 2014 (hereafter RIK14) applied systematically and for the first time \pf{} on the BHB MAXI J1543-564 fitting single hump power spectra in a single energy band. RIKK16 then applied \pf{} on the BHB MAXI J1659-152 using for the first time the hypothesis of fluctuations stirred up in and propagating from the disc. They fitted simultaneously power spectra in two energy bands and cross-spectra between these two bands. RIK17 further updated \pf{} introducing the hypothesis of extra variability in the hot flow, damping, and different propagation speeds of the fluctuations. Veledina (2016) obtained multi-hump power spectra considering a slightly different model. The multi-hump shape of the noise is the result of interference between two variable spectral components: Compton up-scattered disc photons and synchrotron, hot flow, photons. The variability is triggered by mass accretion rate fluctuations propagating from the disc through a hot flow. \\
RIK17 used \pf{} to study two observations of the BH XTE J1550-564. Their analysis showed qualitative as well as quantitative differences between data and model predictions. These discrepancies represent an important challenge for the propagating fluctuations scenario. RIK17 speculated that the process generating the QPO may also influence the broad band noise. In this case, the observed variability is not only due to propagating fluctuations and any \pf{} fit of the power spectrum in two energy bands and of the phase lag between these two bands, would be biased. Here, in order to further explore this possibility and, more in general, the validity of the propagating fluctuations model, we apply \pf{} to selected Cygnus X-1 observations (that in general do not show QPOs). Because we want to test \pf{} on the broadest variety of observational states and to compare our results with previous analyses, we used the same soft state observation as that selected by AU06, and a hard state observation showing an energy spectrum that is very similar to one of the XTE J1550-564 observations analyzed by RIK17. We also selected a third observation in the transition from soft to hard state. \\
In Sec. \ref{sec:cyg} we report the main characteristics of Cygnus X-1 relevant to this study and in Sec. \ref{sec:pro} we briefly summarize the \pf{} model features. Sec. \ref{sec:obs} and Sec. \ref{sec:res} describe data reduction and our results, respectively, and in Sec. \ref{sec:dis} we discuss the results. We conclude that \pf{} predictions are consistent with observations during the soft and hard state, while the propagating fluctuations scenario cannot explain the variability characteristics observed during the intermediate state.

\section{Cygnus X-1}
\label{sec:cyg}

Cygnus X-1 (Bowyer et al. 1965) is a bright ($\sim$ 0.6 Crab) persistent BHB. Up to date, it has been mostly observed in the hard state with frequent (and often very fast) transitions to the soft state (Grinberg et al. 2013). The Cygnus X-1 energy spectrum in the hard state is characterized by a non-thermal component, usually modelled with a power-law ($\Gamma \approx$  1.6) with an exponential cut-off above $\approx$ 100 keV, and a thermal component, usually modelled with a multi-temperature blackbody (Dove et al. 1998 and references therein). Compared to the canonical LHS, the Cygnus X-1 hard state differs in the lack of QPOs. The Cygnus X-1 soft state also differs from the High-Soft State (HSS) of the canonical phenomenological description of transient BHBs (e.g. Belloni 2010; Remillard \& McClintock 2006). The HSS energy spectrum is dominated by thermal emission (usually modelled with a multi-temperature black body), high luminosity, and low amplitude rapid variability (fractional rms $\approx$ 3\%), while in Cygnus X-1 we additionally observe a significant non-thermal power law emission and high rapid variability (fractional rms $\approx$ 25\%).\\
In a Power x Frequency representation (the one adopted in this study), the power spectrum of the source in the hard state shows two broad humps, while in the soft state it shows a single, continuous, broad band component that can be modeled with a single power law  (Axelsson et al. 2005). In general, Cygnus X-1 power spectra do not show strong QPOs, even though there is evidence of the presence of broad QPOs in many observations (Belloni et al. 1996; Cui et al. 1997).\\
The drastic differences associated with different states in the spectral and timing properties of Cygnus X-1 suggest very different accretion regimes in the hard and soft state. Many models have been proposed to explain the behaviour of the source, but all of them agree on the presence of two main physical components (e.g. Zdziarski et al. 2002): a geometrically thin optically thick disc (producing the thermal radiation) and an optically thin hot inner flow or corona (producing the non-thermal/Comptonized radiation). The exact geometry of the system, and the exact way these two components interact with each other in the various states, is still a matter of debate.\\
On 10 May 1996, Cygnus X-1 started a transition from the hard to the soft state. The source remained in the soft state for about 2 months before going back to the hard state (Cui et al. 1997). Studying the variability in different energy bands, Churazov et al. (2001) found that this soft state is characterized by a soft stable multi-temperature blackbody component and a harder variable power-law component. The physical scenario suggested by these authors is an optically thin variable corona `sandwiching'' an optically thick stable disc. In this scenario mass accretion rate fluctuations originate in the corona and propagate towards the BH, producing the observed rapid variability. AU06 selected one observation in this soft state to test a Monte Carlo implementation of the mass accretion rate fluctuations model (Lyubarskii 1976; Kotov et al. 2001). They found good agreement between model predictions and data.\\

\section{The PROPFLUC model (summary)}
\label{sec:pro}

\begin{figure} 
\center
\includegraphics[scale=0.4,angle=0,trim=0cm 0cm 0.5cm 0,clip]{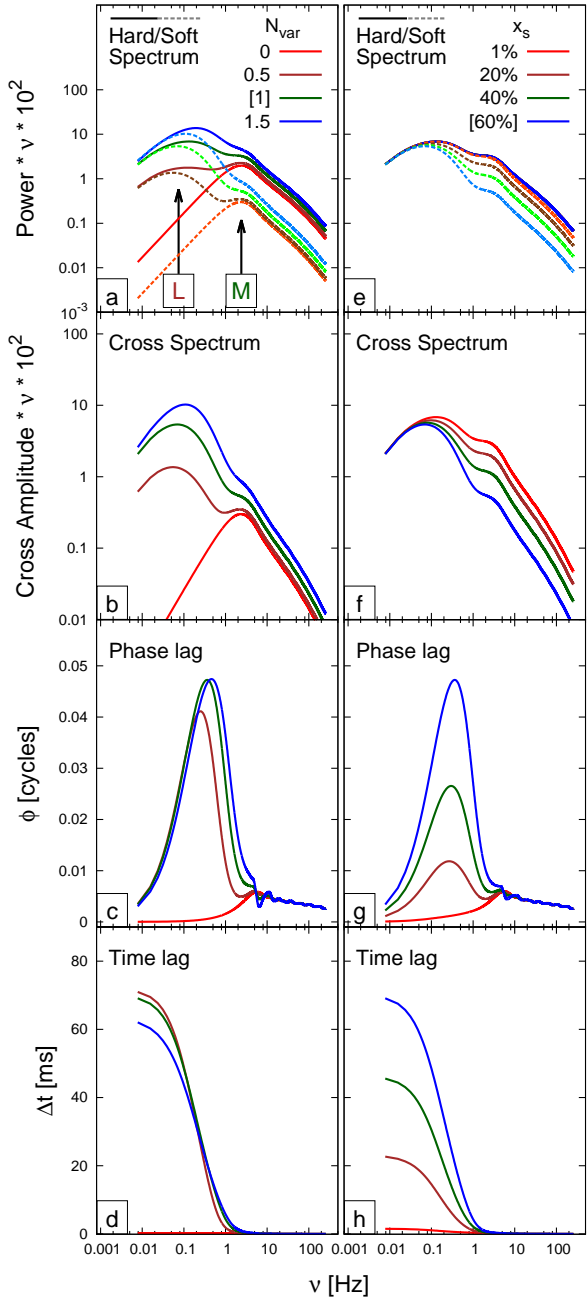}
\caption{Soft (dashed line) and hard (solid line) double hump power spectra, cross spectra, phase lags, and time lags computed varying the disc variability $N_{var}$ and disc fraction in the soft band $x_s$. The double hump power spectra consist of a low-frequency and main hump (L-M configuration). Numbers in square brackets indicate the parameter value for all the other computations.}
\label{fig:comp}
\end{figure}
\pf{} (ID12, IK13) assumes a truncated disc/hot flow geometry: an optically thick geometrically thin accretion disc is truncated at a certain radius \ro{}. Inside this radius, accretion takes place through an optically thin (opacity $\tau \sim$ 1) geometrically thick hot flow. Mass accretion rate fluctuations are stirred up at each radius of both the disc and the hot flow. The fluctuations propagate towards the BH (inward-only propagation) throughout the disc/hot flow system. Simultaneously to the propagation of fluctuations, if the accretion flow is misaligned with the equatorial plane of the BH, the entire hot flow (but not the disc) experiences Lense-Thirring (LT) precession because of frame dragging close to the BH. This precession produces a QPO signal. \\
The fluctuations propagate towards the BH on the local viscous time scale. A further assumption is that the characteristic time scale of the fluctuations is (also) set by the local viscous time scale, so that the characteristic frequency of the fluctuations is the local viscous frequency. \pf{} assumes that, in the disc, the radial profile of the viscous frequency follows the Shakura \& Sunyaev (1973) prescription. In particular $\nu_{v,disc}(r) = \nu_{d,max}  (r/r_o)^{-3/2}$ (Shakura \& Sunyaev 1973), where lowercase $r$ is the radial coordinate scaled by the gravitational radius $r = R/{R_\text{g}}$ ($R_g = GM/c^2$) and $\nu_{d,max} = \nu_{v,disc}(r_o)$ is a model parameter (RIKK16). In the hot flow, the viscous frequency is described by a smoothly-broken power law bending at the bending wave radius \rbw{} (ID12, IK13, RIK14, RIKK16):
\begin{equation}
\nu_{v,flow} (x) = \frac{1}{2\pi r_\text{bw}^2 \Sigma_0} \frac{(1+x^\kappa)^{(\zeta+\lambda)/\kappa}}{x^{\lambda+2}}\frac{c}{R_\text{g}}
\label{eq:fv}
\end{equation}
where $x = r/r_\text{rb}$. In this expression $\Sigma_0$, $r_\text{bw}$, and the indices $\kappa$, $\zeta$, and $\lambda$ are all model parameters. The bending wave radius depends on the scale-height factor (H/R) of the hot flow through the expression $r_\text{bw} = 3 (H/R)^{-4/5} a_*^{2/5}$ (ID12), where $a_*$ is the dimensionless spin parameter of the BH and it is a model parameter (see table 1 in RIK17 for a detailed description of the model parameters).\\
In \pf{}, at each instant in time, the total flux in a certain energy band is a linear combination of fluctuating mass accretion rate at each radius (IK13). The coefficients of this linear combination are set by the emissivity radial profile and depend on the emission mechanisms of the two accreting regions. \pf{} assumes multi-temperature blackbody emission in the disc (following the Shakura \& Sunyaev 1973 prescription) and a power-law radial emissivity profile in the hot flow (ID12, IK13). In this way, mass accretion rate fluctuations are converted into X-ray flux variability and, because of the propagation, the variability produced at a certain radius is characterized by both fluctuations on time scales around the local viscous time scale and by longer time scale fluctuations propagated from larger radii. \\
Because fluctuations propagate in a finite time (the viscous time scale), \pf{} naturally predicts positive (hard) phase lags between a soft and a hard energy band. The amplitude (and the sign) of the lags depends on the radial dependence of the energy spectrum emitted by the accretion flow and on the selected energy bands (RIKK16, RIK17). The amplitude of the fluctuations is also expected to be a function of radius. In the disc, this dependence is assumed to be a Gaussian peaking at the boundary between disc and hot flow, the truncation radius \ro{} (RIKK16). In this way, only the disc portion close to the hot flow contributes significantly to the observed variability. In the hot flow, the fluctuation amplitude can be either constant (ID12, IK13, RIKK16), or a constant plus a narrow Gaussian peaking at the bending wave radius, \rbw{} (RIK17). In the latter case, we talk about extra variability in the hot flow. This assumption is made to take into account results from numerical simulations (Fragile \& Blaes 2008; Henisey, Blaes \& Fragile 2012), showing that the inner part of a tilted accretion flow may produce extra high-frequency variability. \\
The viscous frequency radial profile, the emissivity radial profile, and the amount of variability injected at each radius of the disc/hot flow system, together determine the shape of the power and cross-spectrum between two energy bands (AU06, RIKK16, RIK17). Constant amplitude variability stirred up and propagating only in the hot flow produces a single hump power spectrum (M configuration, RIK17). In the P$\nu$ vs $\nu$ representation, a single hump power spectrum consists in a single broad feature extending from a low- to a high-frequency break. These two breaks are set by the viscous frequency at the largest and smallest radius, respectively. The width of the single hump then depends on the extension of the propagating region. In particular, if the propagating region extends to very large radii (so that the low-frequency break is smaller than the frequency resolution), the single hump power spectrum would be a continuous $1/f$ noise feature up to the high-frequency break. We also obtain a single hump power spectrum when we consider variability only from the disc (L configuration). Combining variability from the disc and from the hot flow, we obtain a double hump power spectrum (L-M configuration, RIKK16, RIK17). Considering also extra variability in the hot flow, we obtain a three-hump power spectrum (L-M-H configuration, RIK17). For all the configurations, the emissivity profiles in the soft and hard energy band affect the shape of the power spectrum and the lags associated to the different humps. For example, Fig. \ref{fig:comp} shows double hump (L-M configuration) power spectra, cross-spectral amplitude, and phase lags between a soft and a hard energy band. For completeness, we also plot the time lag $\Delta t = \phi / 2 \pi \nu$. We obtained the different curves varying the amount of variability propagating from the disc ($N_{var}$, Fig. \ref{fig:comp}$a-d$) and the fraction of soft band photons emitted from the disc ($x_s$, Fig. \ref{fig:comp}$e-h$). When the disc variability tends to zero ($N_{var}$ = 0), the double hump converges to a single hump power spectrum (Fig. \ref{fig:comp}$a-d$, red line). When the disc emissivity is very small ($x_s = 1\%$), soft and hard power spectrum are almost identical and the phase lag amplitude is close to zero (Fig. \ref{fig:comp}$e-g$, red line). When $N_{var}$ =1 and $x_s$ varies between 20 and 60\% (Fig. \ref{fig:comp}$e-h$), we obtain different phase lag and soft power spectrum profiles without modifying the shape of the hard power spectrum. The fringes observed at high frequency in the phase lag profile (see Fig. \ref{fig:comp}$c$ and $g$ above $\approx$ 8 Hz) are the result of interference between contributions from different rings.\\
The \pf{} parameters regulate the viscous frequency profile, the amplitude of the variability, and the emissivity profiles in the disc/hot flow. Tab. 1 in RIK17 shows a brief description of the \pf{} parameters. From spectral fitting and flux measurements it is possible to estimate the maximum temperature in the disc $T_{max}$ and the disc fraction in the soft band $x_s$. These two parameters and the hardness ratio HR together set the emissivity of the disc. The model also accounts for damping of the fluctuations as they propagate and for the propagation speed diverging from the value predicted by the local viscous frequency (parameters $D$ and $x_{lag}$, respectively, see RIK17). Because \pf{} does not include a physical model for QPO modulation, the QPO is assumed to be a Lorentzian. The frequency of the QPO is a weighted radial average of the point particle Lense-Thirring precession frequencies in the hot flow (Liu \& Melia 2002; ID12, eq. 1). The QPO characteristics (amplitude, coherence, and phase lag) are \emph{ad hoc} parameters. \\

\section{Observations and data analysis}
\label{sec:obs}

\begin{figure} 
\center
\includegraphics[scale=0.55,angle=0]{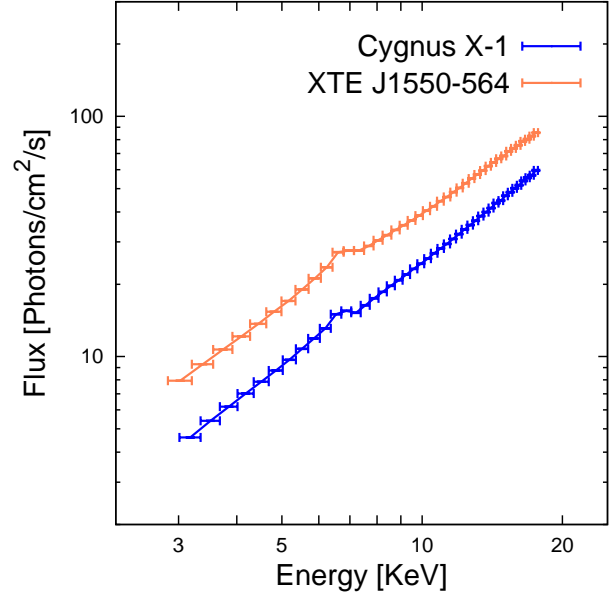}
\caption{Unfolded energy spectrum of Cygnus X-1 (blue line) and XTE J1550-564 (orange line) in the LHS. These two observations are characterized by a very similar energy spectrum.}
\label{fig:compspe}
\end{figure}
We analyzed data from the Proportional Counter Array (PCA; Jahoda et al. 1996) on board of the \textit{Rossi X-ray Timing Explorer} (RXTE). We used 3 pointed observations: 1) 2 February 1997 (obs. ID 10238-01-03-00, MJD 50482, hard state), 2) 12 August 1996 (obs. ID 10412-01-07-00, MJD 50307, transition from soft to hard state), and 3) 18 June 1996 (obs. ID 10512-01-09-01, MJD 50252, soft state). The three observations contain $\sim$ 3.5, 2.1 and 2.4 ks of data, respectively.\\
We performed spectral analysis using HEASOFT 6.13. We used Standard2 data (16s time resolution) to extract source and background spectra. For the three analyzed spectra, we created a PCA response matrix, we corrected the energy spectra for background, and we applied a systematic error of 1\%. We fitted the energy spectra in the 3-20 keV energy band using XSPEC 12.8.2 (Arnaud 1996) with the abundances of Wilms et al. (2000). \\
As noted by Wilson \& Done (2001), the energy spectrum of the first observation is very similar to the spectrum of XTE J1550-564 in the LHS at the beginning of the 1998-1999 outburst. Specifically, there is an evident similarity between the energy spectrum of the first observation and the observation 30188-06-01-01 of XTE J1550-564 (see Fig. \ref{fig:compspe}). The latter was selected by RIK17 for testing \pf{}. For comparison purposes, we used the same spectral model as that adopted by RIK17 (following Axelsson et al. 2013) for fitting the first observation: \textsc{TBABSxGABSx(DISKBB+NTHCOMP+RFXCONVxNTHCOMP)} (Mitsuda et al. 1984; Zdiarski et al. 1996; $\rm \dot{Z}$ycki et al. 1999; Kolehmainen et al. 2001). We varied the \textsc{diskbb} normalization, the \textsc{nthcomp} spectral index and normalization, the amount of reflection R, we used an inclination of 50\degree (Sowers et al. 1998), and we fixed all the other model parameters to the best-fit parameters found by Axelsson et al. (2013). We obtained a $\chi^2_{red}$ = 0.95 with 38 degrees of freedom. For the second and third observation, we used the model \textsc{TBABSx(GAUSSIAN+DISKBB+NTHCOMP)} (Mitsuda et al. 1984; $\rm \dot{Z}$ycki et al. 1999) obtaining a $\chi^2_{red}$ = 0.96 and 0.45 with 36 degrees of freedom, respectively. From our spectral fits we obtained the disc contribution to the soft band $x_s$ and the maximum temperature in the disc $T_{d,max}$ for each of the selected observations (see Tab. \ref{tab:specfit}). We emphasize that the goal of our spectral fitting is estimating $T_{d,max}$ and $x_s$ (two parameters necessary for \pf{} fits), and not to perform a detailed spectral analysis.\\
For all our observations, we selected the same two energy bands as RIK17: 1.9 - 13.0 keV (soft) and 13.4-20.3 keV (hard). Source and background light curves were extracted from these bands. We computed the count ratio between hard and soft band (i.e. the hardness ratio $HR$, another information necessary for \pf{} fitting) from the background subtracted light curves of each observation.\\
\begin{table}
\setlength\extrarowheight{4pt}
\renewcommand{\arraystretch}{1.5}
\centering
\caption{Best-fit $T_{d,max}$, $x_s$, reduced $\chi^2$, and degrees of freedom obtained from spectral fitting for the three observations we analyzed.}
\label{tab:specfit}
\begin{tabular}{|c|c|c|c|}
\hline
Observation & $T_{d,max}$ [keV]& $x_s$[\%] & $\chi_{red}^2$ / d.o.f.\\
\hline 
\hline 
10238-01-03-00 (1) & 0.55 (fixed) & 2.9$_{-0.2}^{+0.2}$ & 0.95 / 38\\
10412-01-07-00 (2) & 0.58$_{-0.03}^{+0.03}$ & 66.6$_{-0.9}^{+4.0}$ & 0.96 / 36\\
10512-01-09-01 (3) & 0.58$_{-0.05}^{+0.05}$ & 61.8$_{-1.2}^{+8.7}$ & 0.45 / 36\\
\hline
\end{tabular}
\end{table}
For all the observations, we combined Single Bit mode (time resolution $\approx$ 125 $\mu$s) and Event mode (time resolution $\approx$ 62 and 16 $\mu$s for observations 1-2 and 3, respectively) to perform Fourier timing analysis. Leahy-normalized power spectra were computed in the soft and hard band using a time resolution of 1/8192 s and 128 s data segments. This gives a frequency resolution of 1/128 $\approx$ 0.008 Hz and a Nyquist frequency of 4096 Hz. Using the same setting, we computed cross-spectra between soft and hard band. We averaged power and cross-spectra, subtracted the Poisson noise, and applied fractional rms normalization following RIKK16. \\

\section{Results}
\label{sec:res}

\begin{figure} 
\center
\includegraphics[scale=0.7,angle=0,trim=0.cm 0cm 0.cm 0cm,clip]{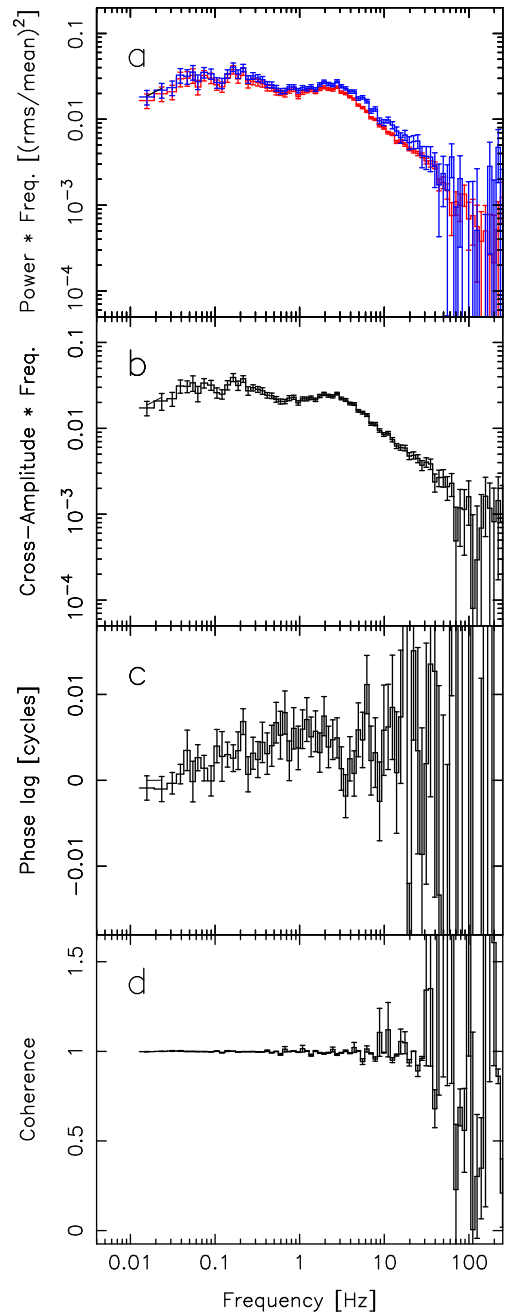}
\caption{Hard (blue line) and soft (red line) power spectrum (a), cross-spectrum amplitude (b), phase lag (c), and intrinsic coherence (d) of the first observation (10238-01-03-00).}
\label{fig:obs1}
\end{figure}
\begin{figure} 
\center
\includegraphics[scale=0.7,angle=0,trim=0.cm 0cm 0.cm 0cm,clip]{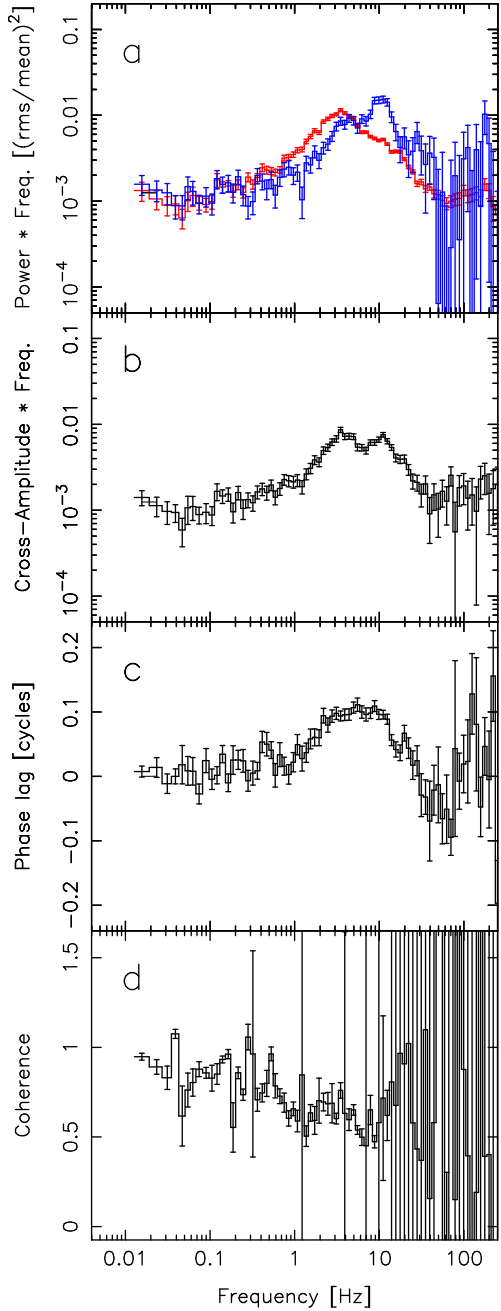}
\caption{Hard (blue line) and soft (red line) power power spectrum (a), cross-spectrum amplitude (b), phase lag (c), and intrinsic coherence (d) of the second observation (10412-01-07-00).}
\label{fig:obs2}
\end{figure}
\begin{figure} 
\center
\includegraphics[scale=0.7,angle=0,trim=0.cm 0cm 0.cm 0cm,clip]{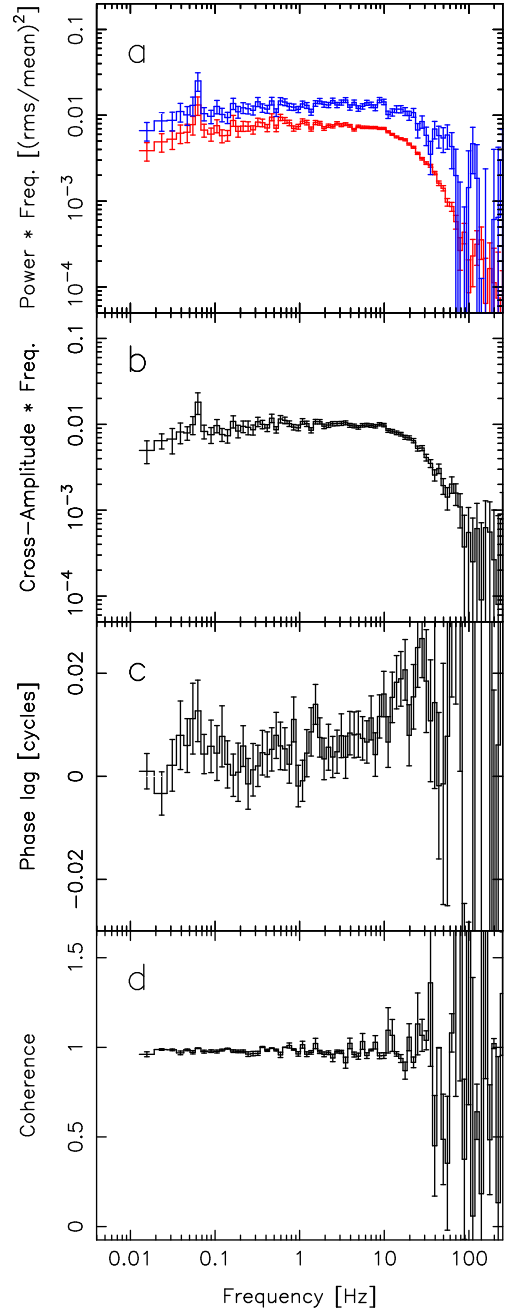}
\caption{Hard (blue line) and soft (red line) power spectrum (a), cross-spectrum amplitude(b), phase lag (c), and intrinsic coherence (d) of the third observation (10512-01-09-01).}
\label{fig:obs3}
\end{figure}
\begin{figure}
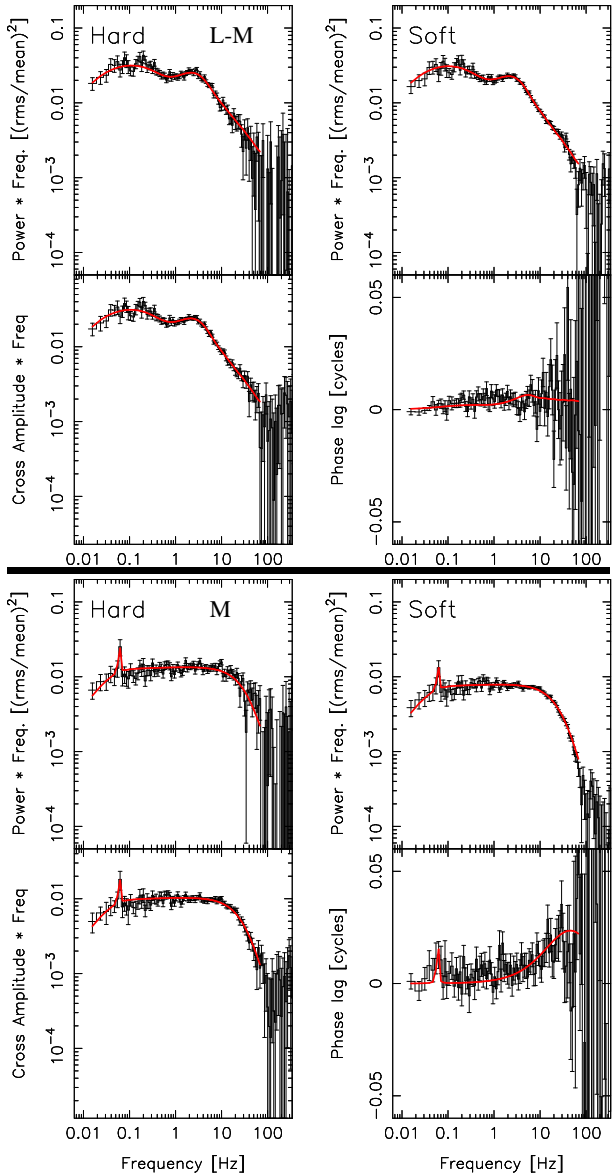

\center
\includegraphics[scale=0.45,angle=0,trim=0cm 1.cm 0 0cm,clip]{figs/obs3_LM.ps}
\vspace{-0.5cm}
\rule[0.525cm]{8cm}{2.pt}
\includegraphics[scale=0.45,angle=0,trim=0cm 0.cm 0 0,clip]{figs/obs1_M.ps}
\caption{Best fit of the first (top) and third (bottom) observation (solid red line) different hump configurations: only main (M) and low-frequency and main (L-M).}
\label{fig:fit}
\end{figure}
We jointly fitted logarithmically binned soft and hard power spectra, and cross-spectra between the two bands with the propagating mass accretion rate fluctuations model \pf{} (IK13, RIKK16, RIK17). We used the same resolution for data and model and we fitted data up to 70 Hz. For the first (hard state) and second (intermediate state) observation, we used a double hump power spectrum (L-M configuration). This is obtained combining mass accretion rate fluctuations stirred up in and propagating through the disc and the hot flow. In this configuration, the physical properties of the accretion flow vary sharply at the truncation radius \ro{} and this leads to two different humps in the power spectrum (see Sec. \ref{sec:pro}). For the third (soft state) observation, we fitted the data using only a main hump power spectrum. This can be obtained considering mass accretion rate fluctuations generated in and propagating through a single region (so without abrupt variations in viscous frequency, emissivity radial profile, etc). This single region can be either the hot flow (M configuration) or the disc (L configuration). For all the fits we considered a 10 M$_{\odot}$ BH with dimensionless spin parameter $a_* = 0.5$. We used $N_{dec}$ = 35 rings per radial decade, and the fixed hydrogen column density ($n_H$ = 0.6 $\times$ 10$^{22}$ cm$^{-2}$, Balucinska-Church et al. 1995).\\
Fig. \ref{fig:obs1}-\ref{fig:obs3} show soft and hard power spectrum ($a$), cross-spectral amplitude ($b$), phase lag ($c$), and coherence ($d$) between these two bands for the three observations we analyzed. Fig. \ref{fig:fit} shows the \pf{} best fit of the first and third observation. The best-fit parameters are listed in Tab. \ref{tab:fit}.

\subsection{First (hard state) observation: 10238-01-03-00}
This hard state observation is characterized by a double hump power spectrum: a low-frequency hump peaking at $\approx$ 0.1 Hz and a high-frequency hump peaking at $\approx$ 3 Hz (Fig. \ref{fig:obs1}$a$). Hard and soft power have a very similar shape, even if the hard fractional variability is slightly larger at higher frequencies (above $\approx$ 1 Hz). The maximum amplitude of the lags is $\approx$ 0.005 cycles at $\approx$ 1 Hz (Fig. \ref{fig:obs1}$c$) and the coherence is $\approx$ 1 up to $\approx$ 10 Hz (Fig. \ref{fig:obs1}$d$). \\
We fitted the observation using the L-M configuration (low-frequency and main hump) obtaining a reasonable fit ($\chi^2/dof$ =  444.80/350 $\approx$ 1.3, Fig. \ref{fig:fit} and Tab. \ref{tab:fit}). Because of the small disc contribution to the soft band ($x_s$ $\approx$ 4 \%), the L-M configuration predicts very small phase lags, consistent with the data (Fig. \ref{fig:fit}).
 
\subsection{Second (intermediate) observation: 10412-01-07-00}
This observation covers part of the transition from soft to hard state in 1996 (Cui et al. 1997). Up to $\approx$ 0.5 Hz, soft and hard power show similar shape and normalization (Fig. \ref{fig:obs2}$a$). Above this frequency, soft and hard power show evident differences. In particular, the soft power is characterized by a single hump peaking at $\approx$ 2 Hz. The hard power shows a narrow feature (QPO) at $\approx$ 10 Hz on top of a single hump. The single hump in the hard band peaks at higher frequency than in the soft band ($\approx$ 3-5 Hz). In the soft band we do not observe any QPO. The phase lag spectrum is characterized by a ``bump'' peaking at $\approx$ 6 Hz with amplitude $\approx$ 0.1 cycles (20 times higher than in the previous observation, see Fig. \ref{fig:obs2}$c$). The coherence also clearly differs from the previous observation (Fig. \ref{fig:obs2}$d$): it decreases to $\approx$ 0.5 around 7 Hz.\\
We attempted to fit this observation using the L-M configuration, but we did not get a statistically acceptable result. More specifically, the model does not predict the observed bump in the phase lags ($\approx$ 0.1 cycles). Because of the complex shape of the power spectrum (multiple hump power spectrum), we could not obtain an acceptable fit with either the M or L configuration.

\subsection{Third (soft state) observation: 10512-01-09-01}
\begin{figure} 
\center
\includegraphics[scale=0.4,angle=0,trim=0cm 0cm 0.5cm 0,clip]{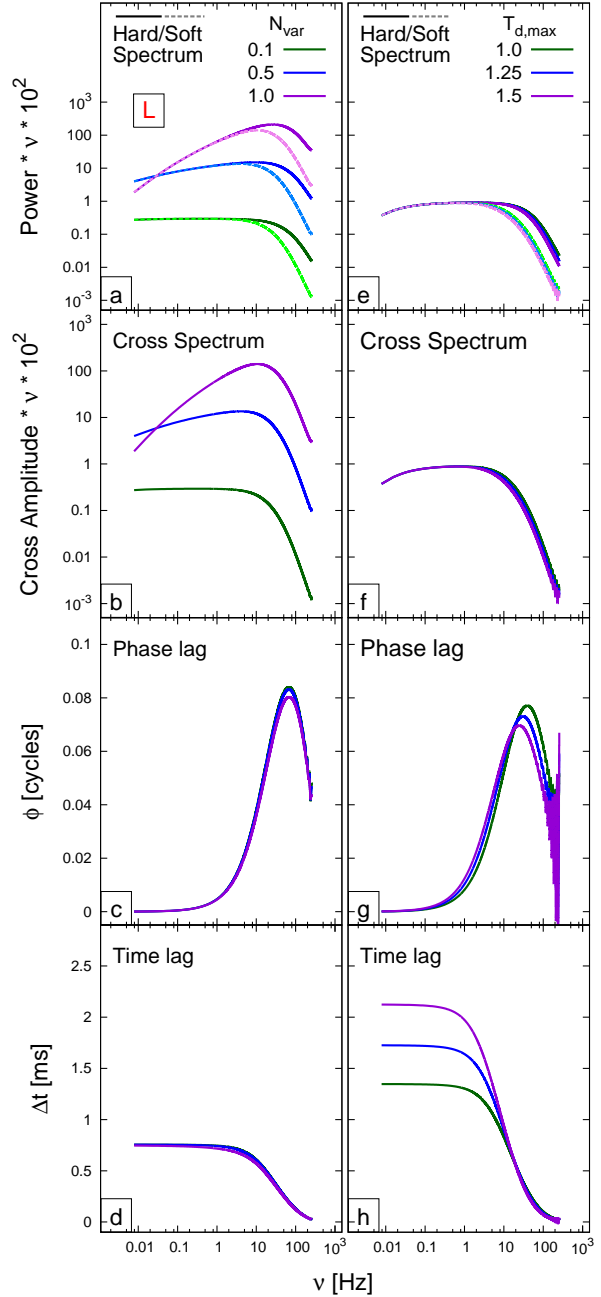}
\caption{Soft (dashed line) and hard (solid line) single hump power spectrum, cross spectra, phase lags, and time lags computed varying the disc variability $N_{var}$ and the maximum temperature in the disc $T_{d,max}$. In these computations, the single hump power spectrum is due to mass accretion rate fluctuations stirred up in and propagating only in the disc (L configuration).}
\label{fig:compd}
\end{figure}
This soft state observation is characterized by $\sim$ 1/$f$ continuous noise (flicker noise, flat shape in the $\nu P(\nu)$ representation) between $\approx$ 0.1 and 10 Hz in both the soft and the hard energy band (Fig. \ref{fig:obs3}$a$). Above 10 Hz, the power decreases steeply with frequency. Below 0.1 Hz the power decreases slightly in both the energy bands. In general, hard and soft power have a very similar shape, but different normalizations, with more fractional variability in the hard band at all frequencies. The amplitude of the lags gradually increases with frequency (Fig. \ref{fig:obs3}$c$), up to $\approx$ 0.02 cycles above 1 Hz. The coherence is $\approx$ 1 up to $\approx$ 10 Hz (Fig. \ref{fig:obs3}$d$). This observation also shows a highly coherent, and significant ($> 3\sigma$), QPO at $\approx$ 0.06 Hz (quality factor Q $\approx$ 38 in both hard and soft band). This feature is characterized by a hard lag of 0.031$^{+0.010}_{-0.003}$ [cycles]. We note that the predicted phase lag at $\approx$ 0.06 Hz (Fig. \ref{fig:fit}, red line) is slightly smaller than the best-fit value of the QPO lag. This difference is because, in the \pf{} prescription, the QPO signal is added to the broad band noise and this noise dilutes the phase lag purely due to the QPO signal. Since the QPO does not follow any of the defining properties of canonical low-frequency QPO classifications (e.g. Casella et al. 2005), we do not tie its frequency to the LT precession frequency, choosing instead to simply model the feature with a Lorentzian.\\
We fitted this observation using both the L and M configuration (single hump power spectrum). Previous analysis of 1996 soft state observations (Churazov et al. 2001) suggests that the disc does not contribute to the variability. However, we still tried to obtain a fit in the L configuration both to explore the predictions of the model in the L configuration and to independently confirm the results of previous analysis with \pf{}. With the L configuration (only disc variability), we did not obtain an acceptable fit. Even though it is possible to reproduce the shape of the power in one of the two bands in this configuration, we could not reproduce the difference in normalization between hard and soft band. Fig. \ref{fig:compd} generically demonstrates that the L configuration produces hard and soft power spectra with the same normalization below $\approx$ 10 Hz, contrary to what we observe in the data (Fig. \ref{fig:obs3}$a$). Furthermore, the predicted amplitude of the phase lag in this configuration ($\approx$ 0.08 cycles) is larger than the data ($\approx$ 0.02 cycles). The M configuration, with the disc contribution in the soft band $x_s$ set as a free parameter, produced a better result. We obtained a reasonable fit ($\chi^2/dof$ = 462/352 $\approx$ 1.3, Fig. \ref{fig:fit} and Tab. \ref{tab:fit}) with $x_s$ = 23\% (consequently, $x_h$, the fraction of hard photons contributed by the disc, is smaller than 1\%\footnote{Since the observed hardness ratio, $HR$, and the PCA response matrix are model inputs, the fraction of hard band emission contributed by the disc component ($x_h$) is calculated self-consistently within the model from the disc temperature parameter $T_{d,max}$ (i.e. we assume a multi-temperature blackbody disc spectrum, convolve with the instrument response and calculate $x_h$ from $x_s$, $HR$, and the folded disc model in the hard and soft bands). See RIKK16 for futher details}). In this case (like in the L configuration), there is only one varying component (the hot flow), but the hot flow emission in the soft band is diluted by non-variable disc emission.\\ 
The shape of the observed power spectrum can also be reproduced using the L-M configuration. In this configuration, the power spectrum at low frequency constrains the radial extension of the varying disc, $\Delta d$, instead of the truncation radius, \ro{} (as it was in the M configuration). Using a double hump power spectrum and varying $r_o$ between 10 and 2000, we did not obtain a better fit. This is because the model underestimates the power spectrum in the hard band at frequencies corresponding to the low-frequency hump (the one due to disc variability). The model also predicts extra phase lags associated with the low-frequency hump that are not observed in the data.

\section{Discussion}
\label{sec:dis}

We applied the propagating fluctuations model \pf{} to three observations of Cygnus X-1. We jointly fitted power spectra in two energy bands and the cross-spectrum between these bands. The observations were selected in order to test \pf{} on a broad variety of states (hard state, soft state, and transition between the two) and to compare our analysis with previous results from AU06 and RIK16. Our fits were intended to explore the general agreement between propagating fluctuations model predictions and data; for this reason we did not compute the errors on the best-fit parameters. However, encouraged by the statistically reasonable fits we obtained on the first and third observation, we speculate about the possible geometry of the accreting system suggested by our best-fit values. 

\subsection{First observation}
\begin{figure} 
\center
\includegraphics[scale=0.55,angle=0]{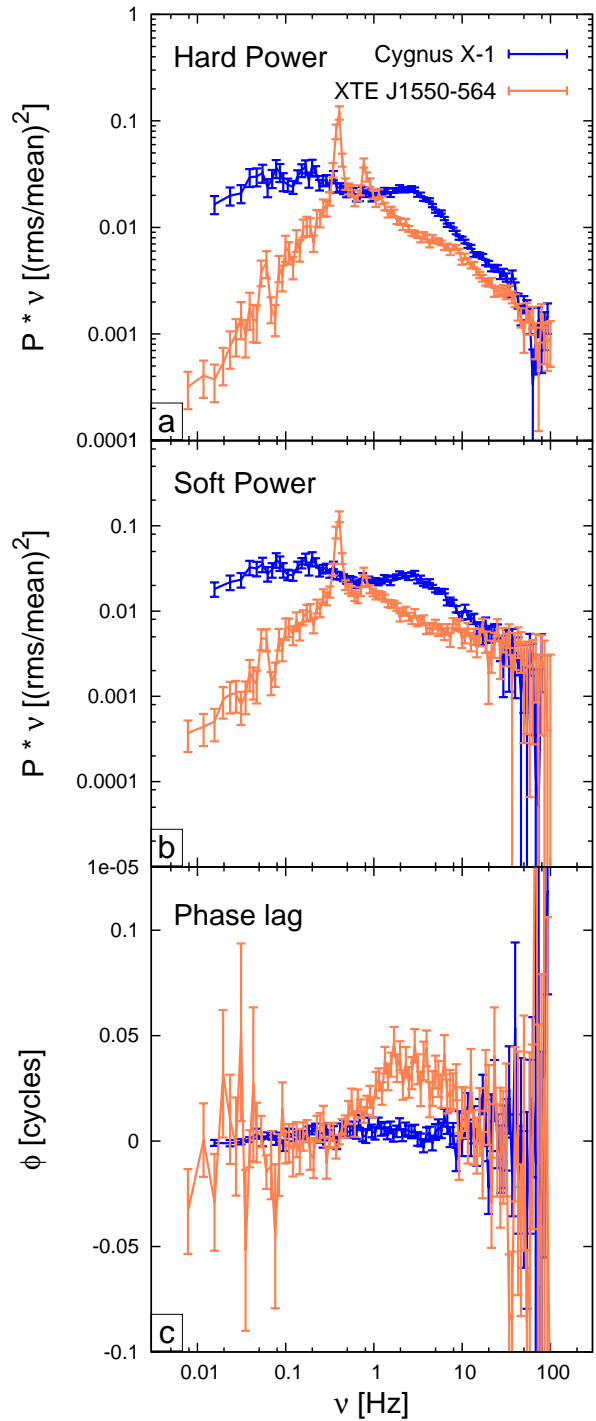}
\caption{Hard power spectrum ($a$), soft power spectrum ($b$), and Phase lag ($c$) between this two energy bands of Cygnus X-1 (blue line) and XTE J1550-564 (orange line) in the hard state. Even if the these two observations are characterized by a very similar energy spectrum, their timing properties are very different.}
\label{fig:comppow}
\end{figure}

In the first observation the source is in the hard state (Gilfanov et al. 1999). We obtained a reasonable fit using the L-M configuration ($\chi^2/dof \approx$ 1.3). This configuration assumes a truncated disc geometry with the hot flow extending radially inside of the disc inner edge. Variability is generated in the region of the disc closest to the truncation radius, and also throughout the inner hot flow (see RIKK16 and RIK17). From our fit, we estimate a truncation radius of $\approx$ 10 \rg{}, with the amplitude of fluctuations generated in the disc peaking at the truncation radius and dropping off following a Gaussian profile with width of $\Delta d \approx$ 52 \rg{}. While it is possible to obtain an acceptable fit in this configuration, we cannot exclude that part of the disc is sandwiched by the hot flow (see Sec. \ref{sec:third}). We did not test this scenario and in all our considerations in this section we assume the simple truncated disc geometry. \\
Our model parameterises the radial surface density profile, and, via mass conservation, the radial dependence of the viscous frequency, with a smoothly broken power-law function. The break radius of this function is assumed to occur at the so-called bending wave radius, which, for a tilted accretion flow, marks the point inside of which the pressure waves that couple the flow (bending waves) cause an oscillatory structure of the tilt angle (Lubow et al. 2002; Fragile et al. 2007; Ingram, Done \& Fragile 2009). This radius is given by $r_{bw} \sim 3 (H/R)^{-4/5} a_*^{2/5}$, where $H/R$ is the scaleheight of the hot flow and $a_*$ is the dimensionless spin parameter of the BH. In our best-fit model, the bending wave radius \rbw{} is smaller than the hot flow inner radius \ri{}, meaning that the viscous frequency is parameterised to a good approximation by a single power-law with no break. Taking this at face value and assuming that the hot flow is indeed tilted as per the Ingram, Done \& Fragile (2009) model, this implies \textit{either} that \rbw{}$<$\ri{} or \rbw{}$>$\ro{}. This is because of the inherent degeneracy in our parameterisation: the single power-law parameterising the viscous frequency in our best fitting model could be the power-law index appropriate for $r >>$\rbw{}, or the index appropriate for $r <<$\rbw{}. From the formula for \rbw{}, and assuming a spin value of $a_*=0.5$, this implies either that $H/R <$ 0.16 or $H/R >$ 0.43. Alternatively, the single power law obtained from the fit could result from the BH spin axis of Cygnus X-1 being aligned with the inner accretion disc and therefore probably with the binary rotation axis. In this case, the model predicts no break in the surface density profile, and also, there would be no LT precession of the hot flow. This could explain why Type-C QPOs (see Casella et al. 2005 for a description of QPO types) are not observed in Cygnus X-1, assuming the Type-C QPOs observed in other sources result from LT precession.\\
We note that Axelsson, Borgonovo \& Larsson (2005) present evidence that Type-C QPOs may in fact be present in Cygnus X-1, but with unusually low rms and quality factor (see fig. 17 therein). Perhaps, therefore, the tilt angle is merely small in this source rather than zero.\\
As already mentioned in Sec. \ref{sec:obs}, the energy spectrum of our observation 1 is very similar to that in the first of the XTE J1550-564 observations analyzed by RIK17 (see Fig. \ref{fig:compspe}). Comparing these two observations, we note that their timing properties are very different (see Fig. \ref{fig:comppow}). The XTE J1550-564 power spectrum is dominated by a strong QPO, while in Cygnus X-1 we do not observe any sharp feature. Both sources show a double hump power spectrum, but the two have rather different characteristics. In XTE J1550-564 both the humps peak at higher frequency, have a smaller amplitude, and a different shape compared to the Cygnus X-1 humps up to $\approx$ 10 Hz. The XTE J1550-564 phase lag profile shows a clear ``bump'' at $\approx$ 5 Hz, while in the third Cygnus X-1 observation the phase lags are very small (Fig. \ref{fig:comppow}$c$). In the propagating fluctuations scenario, the amplitude of the phase lag depends mainly on the emissivity profile in the soft and hard energy bands. These emissivity profiles also affect the shape of the power spectrum, so that similar power spectral shapes in the soft and hard band require similar emissivity profiles in the two bands. If the emissivity indices in two energy bands are equal, the phase lag between these bands is zero. In the XTE J1550-564 observation shown here, the soft and hard power shapes are too similar to produce the observed phase lag ``bump'' at $\approx$ 5 Hz. This discrepancy can be related to either wrong assumptions in the propagating fluctuations model, or some additional physical mechanism producing variability and not considered in \pf{}. In the Cygnus X-1 case, the soft and hard power are also very similar, but the observed phase lags are small enough to be consistent with the model predictions. Since the model seems to work very well for this Cygnus X-1 observation, but very poorly for the XTE J1550-564 observation with a very similar spectrum but with a strong QPO not observed for Cygnus X-1, it seems reasonable to speculate that the misalignment driving the QPO in XTE J1550-564 is also influencing the broad band noise in that source in a manner that our model does not capture. Additional variability stirred up by a mechanism associated with misalignment and not associated with the propagating fluctuations can lead to an incorrect estimate of the emissivities and, consequently, to a different phase lag profile.\\
It is not unreasonable to consider a difference in alignment between BHs in low mass X-ray binaries (LMXBs) and Cygnus X-1, which has a high mass companion. In a LMXB, a BH initially misaligned at birth will align over a timescale comparable to the accretion lifetime (Fragile et al 2001; Martin, Tout \& Pringle 2008; King \& Nixon 2016), suggesting that misalignment is common in LMXBs. The alignment timescale is inversely proportional to the long term mass accretion rate (Fragile et al 2001), which is difficult to ascertain in LMXB BH transients from the few decades of observations of these systems so far. It may have been be higher, and the alignment time scale correspondingly shorter, in Cygnus X-1 which is a persistent X-ray source with accretion driven at least in part by the OB star stellar wind (Gies et al. 2003 and references therein). It is doubtful, however, if the mass accreted during the life time of Cygnus X-1, which is limited by the lifetime of the OB donor star, could have been sufficient to align an initially strongly misaligned BH. Indeed, assuming a mass accretion rate $\dot{\text{M}} \approx 10^{-7} \text{M}_{\odot} yr^{-1}$ and a lifetime of $10^6 yr$, the total mass accreted by the BH in the lifetime of the binary system is M$_{tot}$ $\approx$ 0.1$\text{M}_{\odot}$ (King \& Nixon 2016). For a rotating ($a_*$ = 0.5) BH of 10$\text{M}_{\odot}$, a significant alignment of the BH spin axis with the orbital plane would require a total accreted mass of at least 0.5$\text{M}_{\odot}$ $>$ M$_{tot}$ (King \& Nixon 2016).\\
Another possibility is that the BH spin has always been close to alignment with the binary rotation axis. Asymmetries in the supernova explosion preceding a BH formation can significantly change the orientation of the progenitor spin axis (that is assumed to be equal to the binary rotation axis) and impart a kick to the binary system that affects its proper motion (e.g. Repetto et al. 2012). The proper motion of Cygnus X-1 does not deviate significantly from the motion of the massive star association Cygnus OB3 that it belongs to (Mirabel \& Rodrigues 2003), supporting a formation scenario that does not involve an asymmetric supernova explosion. In this case, the initial BH spin would have already been near-aligned with the binary rotation axis. However, there is also some evidence of misalignment in Cygnus X-1, since reflection spectroscopy implies that the inner disc has a different inclination angle to the binary system (Tomsick et al. 2014). Possibly the exceptional ``focused wind/atmosphere Roche lobe overflow'' accretion geometry in Cygnus X-1 allows for the inner disc to align with the BH spin even when the binary rotation axis is not.

\subsection{Second observation}
The second observation covers part of the transition from soft to hard state. Because of the complex shape of the power spectrum, we attempted to fit this observation with a double hump power spectrum (L-M configuration). For this observation we could not get a statistically acceptable fit with \pf{}. The model does not predict the large phase lags between $\approx$ 0.5 and 11 Hz.\\

\subsection{Third observation}
\label{sec:third}

In the third observation the source is in the soft state (Cui et al. 1997). AU06 successfully applied a Monte Carlo mass accretion rate fluctuations model to this observation. They fitted the ratio between power in a hard and soft band (so they did not predict the shape of the two individual power spectra) and the phase lag between those two bands. In our analysis we fitted simultaneously the soft and hard power spectrum, and also the cross-spectrum (so both phase lags and coherence) between the two bands. We obtained a reasonable fit using the M configuration ($\chi^2/dof \approx$ 1.3), i.e. considering fluctuations stirred up and propagating only in the hot flow. In computing the Fourier products, we considered longer time segments than those used by AU06 (128s as opposed to 32s), extending the frequency range down to $\approx$ 0.01 Hz. The shape of the power spectrum between $\approx$ 0.01 and 0.1 Hz allowed us to constrain the outer edge of the variable region (\ro{} $\approx$ 2500 \rg{}) and to reveal a very low frequency QPO, which does not fit into any of the canonical low-frequency QPO classifications (Casella et al. 2005). The best-fit bending wave radius is again smaller than the inner radius, meaning that the viscous frequency in the hot flow is described by a single power-law, as in our fit to observation 1. \\
\begin{figure} 
\center
\includegraphics[scale=0.7,angle=0,trim=0.cm 0cm 0.cm 0cm,clip]{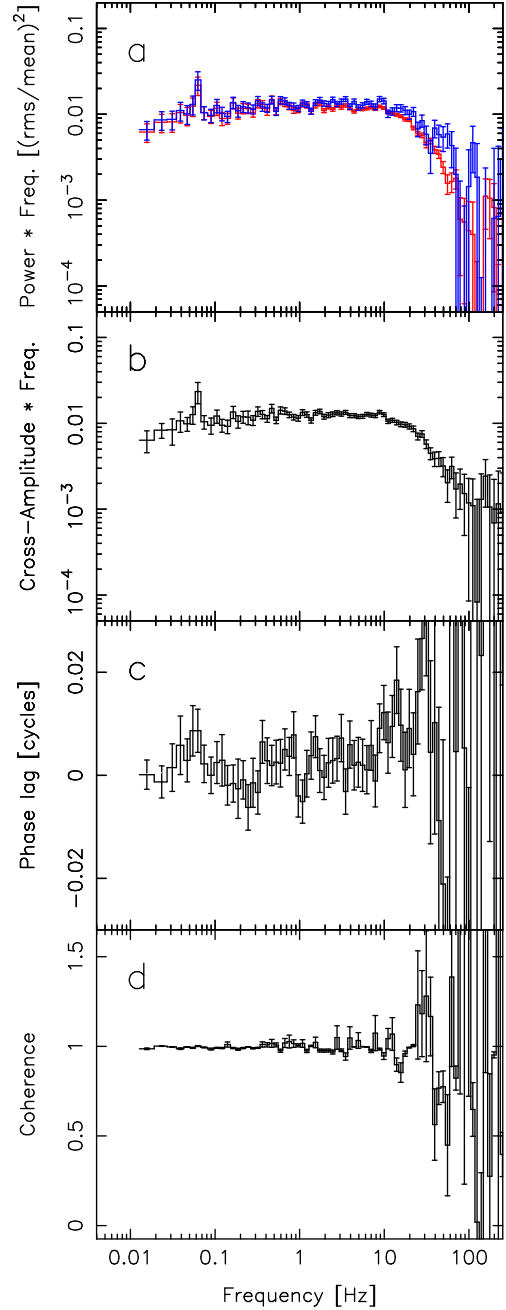}
\caption{Hard (blue line) and soft (red line) power spectrum (a), cross-spectrum amplitude (b), phase lag (c), and intrinsic coherence (d) of the third observation (10512-01-09-01) obtained using a higher energy soft band (6.9-13.0 keV).}
\label{fig:obs1_hard}
\end{figure}
In our fit, no variability is generated in the disc itself ($N_{\rm var}=0$), and the fraction of soft band emission contributed by the disc component ($x_s$) is a free parameter. This allows the model to reproduce the difference in normalization between the observed soft and hard power spectra, whilst preserving their almost identical shapes: the stable disc component dilutes the variable flow component more in the soft band than in the hard band (because $x_s>x_h$). Fig. \ref{fig:compd}$d$ demonstrates that this dilution is required: considering only emission from a single propagating region (either hot flow or disc) leads to identical soft and hard power spectra up to $\approx 10$ Hz, at odds with the observations. Moreover, the fit requires \textit{more} dilution in the soft band than in the hard band; i.e. the stable component is soft. Therefore an alternative model with a variable disc and stable flow will not work. This, together with the fact that we could not obtain an acceptable fit considering purely a variable disc (L configuration), agrees with Churazov et al. (2001), who constrained stable and variable spectral components of Cygnus X-1 for a 1996 soft state observation. They found that the stable component was in the shape of a disc spectrum, which is consistent with what we find here by requiring more dilution in the soft band than in the hard band. We also find that, when we select a higher energy soft band (6.9-13.0 keV, Fig. \ref{fig:obs1_hard}), hard and soft power spectra are very similar to each other. This is consistent with the disc spectrum being constant, since there is no disc emission in this higher energy soft band and therefore no dilution. In this case, fitting the data with \pf{} we obtain a reasonable fit ($\chi^2/dof = 481/351= 1.37$) fixing $x_s$ to the value obtained from spectral fitting ($x_s = 0.52 \%$).\\
Since we measure an enormous outer radius for the flow (\ro{} $\approx 2500 R_g$) but the spectrum clearly requires a strong disc component (requiring a disc inner radius close to the BH, as fitted by Gierlinski et al. 1999), this observation is very much at odds with a truncted disc / hot inner flow geometry. Our fit instead points to a sandwich geometry, in which a stable disc is sandwiched above and below by a variable hot flow / corona, which extends up to large radii and gives rise to the observed rapid variability. This geometry was already suggested by Churazov et al. (2001). Such a geometry implies there must be very strong reflection features in the spectrum, since the covering factor of a sandwich corona is very high. This is consistent with the reflection covering factor of $\Omega/(2\pi)$ =0.7 estimated from the spectral fit of this observation performed by Gierlinski et al. (1999). However, we note that our best-fit value of $x_s$ (23\%) is smaller than the one deduced from spectral fitting ($x_s \approx$ 62\%). This suggests that the disc emission is not entirely stable, with the variability amplitude in the disc following the same radial dependence of variability amplitude as the corona sandwiching it, but with a lower normalisation.

\section{Conclusion}
\label{sec:con}
Using the propagating fluctuations model \pf{} we fitted a soft and a hard state observation of Cygnus X-1. Our analysis suggests a truncated disc geometry in the hard state, with mass accretion rate fluctuations generated in and propagating through both disc and hot flow. The viscous frequency in the hot flow is described by a single power law. This, together with the lack of a type-C QPO, may result because of the alignment of the black hole spin axis with the inner disc and presumably with the binary rotation axis. \\
The energy spectrum of this hard state observation is very similar to a hard state observation of XTE J1550-564 analyzed with \pf{} by RIK17. However, the two observations show very different timing properties, in particular, the XTE J1550-564 power spectrum is dominated by a strong QPO and show an unexplained broad phase lag feature around 5 Hz suggesting that an additional mechanism is at work. RIK17 could not obtain an acceptable fit on this observation with \pf{} and they speculated that the mechanism producing the QPO may also affect the broad band noise in a more complex way than what described in the \pf{} scenario. The fact that in Cygnus X-1 (that does not show type-C QPOs) we obtained an acceptable fit with \pf{}, strengthens this speculation, suggesting that misalignment may be the common cause of both the strong QPO, and the additional variability in XTE J1550-564.\\
We also obtained an acceptable fit for the soft state observation. In this case we considered fluctuations propagating only in the hot flow with the variable emission diluted by a stable (soft) disc emission. This result, together with more detailed spectral analysis from previous studies (Gierlinski et al. 1999), suggests a sandwich geometry, where the variable hot flow sandwiches a stable disc. \pf{} is not designed to properly fit this kind of geometry, however we plan to investigate such configuration in future works. \\
\\
\textbf{ACKNOLEDGEMENTS}\\
The authors thank Andrew King for interesting discussions about black hole spin alignment and the anonymous referee for useful comments that greatly helped to improve the manuscript. The authors acknowledge support from the Netherlands Organization for Scientific Research (NWO).

\begin{table*}
\setlength\extrarowheight{1pt}
\caption{\textsc{propfluc} best fit parameters to the first and third observation. The first observation was fitted using the low-frequency and main hump (L-M) and the third observation using only the main hump (M). The subscripts s and h correspond to soft and hard band respectively. The symbol (f) means that the parameter is fixed. The symbol - means that the parameter was not used in the fit.}
\label{tab:fit}

\begin{tabular}{c|c||c|}

\hline

Observation  & (hard state) 10238-01-03-00 & (soft state) 10512-01-09-01\\
\hline

Humps  & L-M& M\\
\hline\hline

$\Sigma_0$  & 45.43 & 6.48\\
$F_{var} [\%]$ & 89.96 & 41.69\\
$\zeta$ & -  & 1.1\\
$\lambda$ & 5.8 &0.9\\
$\kappa$ & 19.1 &30.0\\
$r_i$ & 4.50 & 4.50\\
$r_{bw}$ & 10.02 $\pm$ 1.67 & 2.87 $\pm$ 1.32\\
$r_o$ & 10.02 & 2489\\
$\gamma_s$ & 5.34 & 9.22\\
$\gamma_h$ & 5.91 & 14.62\\
$(\Delta \gamma)$ & 0.58 & 5.41\\
$N_{var}$ & 0.66 & $-$\\
$\Delta d$ & 51.75 & $-$\\
$\nu_{v,max} [Hz]$ & 0.43 & $-$\\
$T_{d,max}$ [keV] & 0.58(f) & 0.58(f)\\
$x_s$ & 0.03(f)  & 0.23 $\pm$ 0.01\\
$D$ & 1.0 & 1.6\\
$x_{lag}$ & 1.1 & 1.2\\
$M [M_{\odot}]$ & 10.0(f) & 10.0(f)\\
$a_{*}$ & 0.5(f) & 0.5(f)\\
$n_H [10^{22} cm^{-2}]$ & 0.6(f) & 0.6(f)\\
$\chi^2_{red}$ & 1.34 & 1.31\\
$dof$ & 350.00 & 352.00\\
\end{tabular}
\end{table*}

\end{document}